\documentstyle[12pt,thmsa,sw20aip]{article}

\input tcilatex
\QQQ{Language}{
American English
}

\begin{document}

\author{S. T. R. Pinho, T. A. S. Haddad, and S. R. Salinas \\
Instituto de Fisica\\
Universidade de Sao Paulo\\
Caixa Postal 66318\\
05315-970, Sao Paulo, SP, Brazil}
\title{Critical behavior of an Ising model with aperiodic interactions }
\date{14 April 1997 }
\maketitle

\begin{abstract}
We write exact renormalization-group recursion relations for a ferromagnetic
Ising model on the diamond hierarchical lattice with an aperiodic
distribution of exchange interactions according to a class of generalized
two-letter Fibonacci sequences. For small geometric fluctuations, the
critical behavior is unchanged with respect to the uniform case. For large
fluctuations, the uniform fixed point in the parameter space becomes fully
unstable. We analyze some limiting cases, and propose a heuristic criterion
to check the relevance of the fluctuations.
\end{abstract}

\section{Introduction}

The experimental discovery of quasi-crystals motivated many investigations
of the effects of geometric fluctuations produced by different types of
aperiodic structures. There are several specific results for phonon and
electronic spectra of linear problems\cite{ref1}. There are also some
results for more difficult non-linear problems, as the analysis of the
critical properties of a ferromagnetic Ising model on a layered lattice with
aperiodic exchange interactions along an axial direction\cite{ref2}.

In a recent publication\cite{ref3}, Luck presented a detailed study of the
critical behavior in the ground state of the quantum Ising chain in a
transverse field (which is known to be related to the transition at finite
temperatures of the two-dimensional Ising model). The nearest-neighbor
ferromagnetic exchange interactions are chosen according to some
(generalized) Fibonacci sequences. The geometric fluctuations are gauged by
a wandering exponent $\omega $ associated with the eigenvalues of the
substitution matrix of each sequence. The critical behavior remains
unchanged (that is, of Onsager type) for bounded fluctuations (small values
of $\omega $). Large fluctuations induce much weaker singularities, similar
to the case of a disordered Ising ferromagnet.

In this paper, we take advantage of the simplifications introduced by a
hierarchical lattice to give another example of the effects of geometrical
fluctuations on the critical behavior of a ferromagnetic model. It should be
pointed out that hierarchical lattices have been widely used as a toy model
to test critical properties. In some cases the (approximate) Migdal-Kadanoff
renormalization-group transformations on a Bravais lattice are identical to
the (exact) transformations on a suitable hierarchical structure\cite{ref4}.
There are rather detailed studies of these exact transformations for the
Ising model on a variety of hierarchical structures\cite{ref5}. There are
also some recent investigations of an Ising spin-glass on the diamond
hierarchical lattice\cite{ref6}.

This paper is organized as follows. In Section 2, we make some comments on
Fibonacci sequences and introduce a nearest-neighbor ferromagnetic Ising
model on a diamond hierarchical lattice with $q$ branches. To simulate a
layered system, the exchange interactions ($J_A>0$ and $J_B>0$) are
distributed along the branches according to the same aperiodic rule. For a
large class of two-letter Fibonacci sequences, we write exact
renormalization-group recursion relations in terms of two parameters, $%
x_A=\tanh K_A$, and $x_B=\tanh K_B$, where $K_A=\beta J_A$, and $K_B=\beta
J_B$, and $\beta $ is the inverse of the temperature. In Sections 3 and 4,
we analyze two typical and distinct examples. We obtain the fixed points and
the flows of the recursion relations for these specific cases. In the first
example, for small $\omega $, the critical behavior is unchanged with
respect to the uniform case ($J_A=J_B$); the physical fixed point in
parameter space is a saddle point. In the example of Section 4, however, the
fluctuations turn the physical fixed point fully unstable. We also analyze
some limiting cases. For $q=1$ (Ising chain), there is no transition at
finite temperatures. The fixed point at zero temperature, however, may
change its character depending on the value of the wandering exponent. In a
particular infinite-branching limit ($q\rightarrow \infty $), we obtain
rather simple results. In Section 5, we present some conclusions, as well as
a heuristic adaptation for the diamond hierarchical lattice of Luck's
criterion\cite{ref7} for the relevance of geometric fluctuations.

\section{Definition of the model}

Consider a particular two-letter generalized Fibonacci sequence given by the
substitutions 
\begin{equation}
\begin{array}{l}
A\rightarrow AB, \\ 
B\rightarrow AA.
\end{array}
\label{eq1}
\end{equation}
If we start with letter $A$, the successive application of this inflation
rule produces the sequences 
\begin{equation}
A\rightarrow AB\rightarrow ABAA\rightarrow ABAAABAB\rightarrow \cdot \cdot
\cdot .  \label{eq2}
\end{equation}
At each stage of this construction, the numbers $N_A$ and $N_B$, of letters $%
A$ and $B$, can be obtained from the recursion relations 
\begin{equation}
\left( 
\begin{array}{c}
N_A^{\prime } \\ 
N_B^{\prime }
\end{array}
\right) ={\bf M}\left( 
\begin{array}{c}
N_A \\ 
N_B
\end{array}
\right) ,  \label{eq3}
\end{equation}
with the substitution matrix 
\begin{equation}
{\bf M}=\left( 
\begin{array}{cc}
1 & 2 \\ 
1 & 0
\end{array}
\right) .  \label{eq4}
\end{equation}
The eigenvalues of this matrix, $\lambda _1=2$ and $\lambda _2=-1$, govern
most of the geometrical properties. In a more general case (that is, for a
more general rule), at a large order $n$ of the construction, the total
number of letters is given by the asymptotic expression 
\begin{equation}
N^n=N_A^n+N_B^n\sim \lambda _1^n,  \label{eq5}
\end{equation}
where $\lambda _1>\left| \lambda _2\right| $. The smaller eigenvalue governs
the fluctuations with respect to these asymptotic values, 
\begin{equation}
\Delta N^n\sim \Delta N_A^n=\Delta N_B^n\sim \left| \lambda _2\right| ^n.
\label{eq6}
\end{equation}
From these equations, we can write the asymptotic expression 
\begin{equation}
\Delta N\sim N^\omega ,  \label{eq7}
\end{equation}
with the wandering exponent 
\begin{equation}
\omega =\frac{\ln \left| \lambda _2\right| }{\ln \lambda _1}.  \label{eq8}
\end{equation}

Now we consider a nearest-neighbor Ising model, given by the Hamiltonian 
\begin{equation}
{\cal H}=-\dsum\limits_{\left( i,j\right) }J_{i,j}\sigma _i\sigma _j,
\label{eq9}
\end{equation}
with the spin variables $\sigma _i=\pm 1$ on the sites of a diamond
hierarchical structure. In Fig. 1, which is suitable for the period-doubling
Fibonacci rule of Eq.(\ref{eq1}), we draw the first stages of the
construction of a diamond lattice with a basic polygon of four bonds
(ramification $q=2$). As indicated in this figure, we simulate a layered
system by the introduction of the interactions $J_A>0$ and $J_B>0$ along the
branches of the structure. Considering the elementary transformations of
Fig.2, and using the rules of Eq. (\ref{eq1}), it is straightforward to
establish the recursion relations 
\begin{equation}
\tanh K_A^{\prime }=\tanh \left[ 2\tanh {}^{-1}\left( \tanh K_A\tanh
K_B\right) \right] ,  \label{eq10}
\end{equation}
and 
\begin{equation}
\tanh K_B^{\prime }=\tanh \left[ 2\tanh {}^{-1}\left( \tanh {}^2K_A\right)
\right] ,  \label{eq11}
\end{equation}
where $K_{A,B}=\beta J_{A,B}$. In this particular case, these equations can
also be written as 
\begin{equation}
x_A^{\prime }=\frac{2x_Ax_B}{1+x_A^2x_B^2},  \label{eq12}
\end{equation}
and 
\begin{equation}
x_B^{\prime }=\frac{2x_A^2}{1+x_A^4},  \label{eq13}
\end{equation}
where 
\begin{equation}
x_{A,B}=\tanh K_{A,B}.  \label{eq14}
\end{equation}
In Section 3, we analyze the fixed points associated with these simple
recursion relations.

We can use similar procedures to consider much more general Fibonacci rules.
To avoid any changes in the geometry of the hierarchical lattice, in this
paper we restrict the analysis to period-multiplying Fibonacci inflation
rules of two letters (the largest eigenvalue of the substitution matrix, $%
\lambda _1$, gives the multiplication factor of the period). For example,
let us consider the substitutions 
\begin{equation}
\begin{array}{l}
A\rightarrow A^kB^l, \\ 
B\rightarrow A^{k+l},
\end{array}
\label{eq15}
\end{equation}
where $k,l\geq 1$ are two integers. From the substitution matrix, 
\begin{equation}
{\bf M}=\left( 
\begin{array}{ll}
k & k+l \\ 
l & 0
\end{array}
\right) ,  \label{eq16}
\end{equation}
we have the eigenvalues, $\lambda _1=k+l$ and $\lambda _2=-l$, and the
wandering exponent, $\omega =\ln l/\ln \left( k+l\right) $. Now we consider
a diamond lattice with $q$ branches and a basic polygon of $q\left(
k+l\right) $ bonds. The exchange interactions are chosen according to the
general Fibonacci rule of Eq. (\ref{eq15}). We then write the recursion
relations 
\begin{equation}
\tanh K_A^{\prime }=\tanh \left[ q\tanh {}^{-1}\left( \tanh {}^kK_A\tanh
{}^lK_B\right) \right] ,  \label{eq17}
\end{equation}
and 
\begin{equation}
\tanh K_B^{\prime }=\tanh \left[ q\tanh {}^{-1}\left( \tanh
{}^{k+l}K_A\right) \right] .  \label{eq18}
\end{equation}

\section{Irrelevant fluctuations}

Consider again the Fibonacci inflation rules given by Eq. (\ref{eq1}). From
the eigenvalues of the substitution matrix, $\lambda _1=2$ and $\lambda
_2=-1 $, we have $\omega =0$. For the branching number $q=2$, the recursion
relations are given by Eqs. (\ref{eq12}) and (\ref{eq13}). The fixed points
and some orbits of the second iterates of this map are shown in Fig. 3.
Besides the trivial fixed points, there is also a non-trivial fixed point,
given by 
\begin{equation}
x_A^{*}=x_B^{*}=0.543689...,  \label{eq3-1}
\end{equation}
which comes from the solution of the polynomial equation 
\begin{equation}
x_A^8+2x_A^4-4x_A^2+1=0.  \label{eq3-2}
\end{equation}
The linearization about this uniform fixed point yields the asymptotic
expression 
\begin{equation}
\left( 
\begin{array}{l}
\Delta x_A^{\prime } \\ 
\Delta x_B^{\prime }
\end{array}
\right) =C{\bf M}^T\left( 
\begin{array}{l}
\Delta x_A \\ 
\Delta x_B
\end{array}
\right) ,  \label{eq3-3}
\end{equation}
where 
\begin{equation}
{\bf M}^T=\left( 
\begin{array}{ll}
1 & 1 \\ 
2 & 0
\end{array}
\right)  \label{eq3-4}
\end{equation}
is the transpose of the substitution matrix, and 
\begin{equation}
C=\frac{1-\left( x_A^{*}\right) ^4}{2x_A^{*}}=0.839286....  \label{eq3-5}
\end{equation}
The diagonalization of this linear form gives the eigenvalues 
\begin{equation}
\Lambda _1=C\lambda _1=2C=1.678573...,  \label{eq3-6}
\end{equation}
and 
\begin{equation}
\Lambda _2=C\lambda _2=-C=-0.839286....  \label{eq3-7}
\end{equation}
As $\Lambda _1>1$ and $-1<\Lambda _2<0$, the fixed point is a saddle point
with a flipping approximation (in Fig. 3, we draw the trajectories of some
second iterates of this map). Moreover, if we make $J_A=J_B$, it is easy to
see that the same eigenvalue $\Lambda _1$ characterizes the (unstable) fixed
point of the uniform model (see the diagonal flow in Fig. 3). Therefore, in
this particular example the geometric fluctuations are unable to change the
critical behavior of the uniform system.

It is not difficult to check that the same sort of behavior (saddle point;
largest eigenvalue associated with the uniform system) still holds for all
finite values of the branching number $q$ of the diamond structure. In the
limit of infinite branching ($q\rightarrow \infty $, $K_A,K_B\rightarrow 0$,
with $q^2K_A$ and $q^2K_B$ fixed), the recursion relations are particularly
simple, 
\begin{equation}
y_A^{\prime }=y_Ay_B,  \label{eq3-8}
\end{equation}
and 
\begin{equation}
y_B^{\prime }=y_A^2,  \label{eq3-9}
\end{equation}
where $y_A=q^2K_A$ and $y_B=q^2K_B$. The linearization about the uniform
fixed point, $y_A^{*}=y_B^{*}=1$, yields the relations 
\begin{equation}
\left( 
\begin{array}{l}
\Delta y_A^{\prime } \\ 
\Delta y_B^{\prime }
\end{array}
\right) ={\bf M}^T\left( 
\begin{array}{l}
\Delta y_A \\ 
\Delta y_B
\end{array}
\right) ,  \label{eq3-10}
\end{equation}
which correspond to a limiting case of Eq. (\ref{eq3-3}), with $C\rightarrow
1$. As the limiting eigenvalues are given by $\Lambda _1=\lambda _1=2$, and $%
\Lambda _2=\lambda _2=-1$, a linear analysis is not enough to check the
(flipping saddle-point) character of this marginal case.

Another particular case of interest is the simple Ising chain ($q=1$). The
recursion relations are given by 
\begin{equation}
x_A^{\prime }=x_Ax_B,  \label{eq3-11}
\end{equation}
and 
\begin{equation}
x_B^{\prime }=x_A^2,  \label{eq3-12}
\end{equation}
with the same form of Eqs. (\ref{eq3-8}) and (\ref{eq3-9}), but with the
parameters $x_A$ and $x_B$ given by Eq. (\ref{eq14}). As shown in Fig. 4(a),
the zero-temperature fixed point displays the character of a saddle point.
As there is no phase transition at finite temperatures, it cannot be reached
from physically acceptable initial conditions.

\section{Relevant fluctuations}

To give an example of relevant fluctuations, consider the generalized
Fibonacci substitutions, 
\begin{equation}
\begin{array}{l}
A\rightarrow ABB, \\ 
B\rightarrow AAA.
\end{array}
\label{eq4-1}
\end{equation}
From the substitution matrix, 
\begin{equation}
{\bf M}=\left( 
\begin{array}{ll}
1 & 3 \\ 
2 & 0
\end{array}
\right) ,  \label{eq4-2}
\end{equation}
we have the eigenvalues, $\lambda _1=3$ and $\lambda _2=-2$, and the
wandering exponent, $\omega =\ln 2/\ln 3=0.630929...$. For a general
branching number $q$, we can use Eqs. (\ref{eq17}) and (\ref{eq18}), to
write the recursion relations 
\begin{equation}
\tanh K_A^{\prime }=\tanh \left[ q\tanh {}^{-1}\left( \tanh K_A\tanh
{}^2K_B\right) \right] ,  \label{eq4-3}
\end{equation}
and 
\begin{equation}
\tanh K_B^{\prime }=\tanh \left[ q\tanh {}^{-1}\left( \tanh {}^3K_A\right)
\right] .  \label{eq4-4}
\end{equation}

For the particular case $q=2$, Eqs. (\ref{eq4-3}) and (\ref{eq4-4}) reduce
to the simple relations 
\begin{equation}
x_A^{\prime }=\frac{2x_Ax_B^2}{1+x_A^2x_B^4},  \label{eq4-5}
\end{equation}
and 
\begin{equation}
x_B^{\prime }=\frac{2x_A^3}{1+x_A^6},  \label{eq4-6}
\end{equation}
where the parameters $x_A$ and $x_B$ are given by Eq. (\ref{eq14}). In Fig.
5, we show the fixed points and some second iterates associated with these
recursion relations. The nontrivial fixed point, given by 
\begin{equation}
x_A^{*}=x_B^{*}=0.786151...,  \label{eq4-7}
\end{equation}
comes from the physical solution of a polynomial equation. The linearization
about this uniform fixed point yields the relations 
\begin{equation}
\left( 
\begin{array}{l}
\Delta x_A^{\prime } \\ 
\Delta x_B^{\prime }
\end{array}
\right) =C{\bf M}^T\left( 
\begin{array}{l}
\Delta x_A \\ 
\Delta x_B
\end{array}
\right) ,  \label{eq4-8}
\end{equation}
where the substitution matrix is given by Eq. (\ref{eq4-2}), and 
\begin{equation}
C=\frac{2\left( x_A^{*}\right) ^2\left[ 1-\left( x_A^{*}\right) ^6\right] }{%
\left[ 1+\left( x_A^{*}\right) ^6\right] ^2}=0.618033....  \label{eq4-9}
\end{equation}
From the diagonalization of this linear form we have the eigenvalues 
\begin{equation}
\Lambda _1=C\lambda _1=1.854101...,  \label{eq4-10}
\end{equation}
and 
\begin{equation}
\Lambda _2=C\lambda _2=-1.236067....  \label{eq4-11}
\end{equation}
The absolute values of $\Lambda _1$ and $\Lambda _2$ larger than unit
indicate the relevance of the geometric fluctuations. The uniform fixed
point is fully unstable (and there should be no transition for $J_A\neq J_B$%
).

Again, it is not difficult to check that the uniform fixed point remains
unstable for all values of the branching number $q$. In particular, for $%
q\rightarrow \infty $, and $K_A,K_B\rightarrow 0$, with $q^{1/2}K_A$ and $%
q^{1/2}K_B$ fixed, we have the limiting recursion relations, 
\begin{equation}
y_A^{\prime }=y_Ay_B^2,  \label{eq4-12}
\end{equation}
and 
\begin{equation}
y_B^{\prime }=y_A^3,  \label{eq4-13}
\end{equation}
where $y_A=q^{1/2}K_A$ and $y_B=q^{1/2}K_B$. The linearization about the
uniform fixed point, $y_A^{*}=y_B^{*}=1$, yields the relations 
\begin{equation}
\left( 
\begin{array}{l}
\Delta y_A^{\prime } \\ 
\Delta y_B^{\prime }
\end{array}
\right) ={\bf M}^T\left( 
\begin{array}{l}
\Delta y_A \\ 
\Delta y_B
\end{array}
\right) ,  \label{eq4-14}
\end{equation}
with the substitution matrix given by Eq. (\ref{eq4-2}). We then have the
limiting eigenvalues, $\Lambda _1=\lambda _1=3$, and $\Lambda _2=\lambda
_2=-2$, which confirm the unstable character of this fixed point.

For the particular case of the linear chain ($q=1$), we have 
\begin{equation}
x_A^{\prime }=x_Ax_B^2,  \label{eq4-15}
\end{equation}
and 
\begin{equation}
x_B^{\prime }=x_A^3.  \label{eq4-16}
\end{equation}
In agreement with the lack of a phase transition, the fixed point at zero
temperature is unstable [see Fig. 4(b)].

\section{Conclusions}

From the analysis of the exact renormalization-group recursion relations
associated with an Ising model on a variety of hierarchical diamond
structures, we show that aperiodic fluctuations of the ferromagnetic
exchange interactions may change the character of a (uniform) fixed point.
In a particular example, with a small wandering exponent, the fluctuations
are irrelevant. In this case, the critical behavior is still characterized
by the same exponents of the corresponding uniform system. In another
example, however, stronger geometric fluctuations turn the physical fixed
point unstable in the parameter space. Even in one dimension, although there
is no phase transition at finite temperatures, we show that the geometric
fluctuations change the stability of the uniform fixed point at zero
temperature.

As in the work of Luck\cite{ref7}, it should be interesting to devise a
general criterion to gauge the influence of the geometric fluctuations on
the critical behavior of the Ising model on the diamond hierarchical
lattice. For a large lattice, with $q$ branches, the total fluctuation $%
\Delta J$ in the exchange interactions should be proportional to $\Delta N$.
Thus we can write the asymptotic relation 
\begin{equation}
\Delta J\sim \Delta N\sim N^\omega \sim L^{q\omega },  \label{eq5-1}
\end{equation}
where $L$ is a measure of the total length. The critical temperature, $T_c$,
should be proportional to the total value of the exchange (that is, to $L^q$%
). We can then define a reduced temperature, $t=\left( T-T_c\right) /T_c$,
whose fluctuations are given by the asymptotic form 
\begin{equation}
\delta t\sim \frac{L^{q\omega }}{L^q}=L^{q\left( \omega -1\right) }.
\label{eq5-2}
\end{equation}
In the neighborhood of the critical point, we have $L\sim \xi \sim t^{-\nu }$%
, where $\xi $ is a correlation length, and $\nu $ is a critical exponent.
Thus, we can write 
\begin{equation}
\frac{\delta t}t\sim \frac{t^{-\nu q\left( \omega -1\right) }}t=t^\phi ,
\label{eq5-3}
\end{equation}
with the exponent 
\begin{equation}
\phi =-\nu q\left( \omega -1\right) -1.  \label{eq5-4}
\end{equation}
If $\phi >0$, the fluctuations are irrelevant. If $\phi <0$, however, the
critical behavior is changed drastically. To calculate a suitable value of $%
\nu $, we consider the largest eigenvalue of the diagonal form about the
physical fixed point, and write the usual renormalization-group expression 
\begin{equation}
\Lambda _1=C\lambda _1=b^{y_1}=b^{1/\nu }.  \label{eq5-5}
\end{equation}
As the largest eigenvalue $\lambda _1$ of the substitution matrix gives the
multiplication factor of the Fibonacci rule, we write 
\begin{equation}
b=\left( \lambda _1\right) ^{1/q}.  \label{eq5-6}
\end{equation}
From Eqs. (\ref{eq5-5}) and (\ref{eq5-6}), we have 
\begin{equation}
\nu =\frac{\ln \lambda _1}{q\ln \left( C\lambda _1\right) }.  \label{eq5-7}
\end{equation}
Inserting into Eq. (\ref{eq5-4}), we finally have 
\begin{equation}
\phi =-\frac{\ln \lambda _1}{\ln \left( C\lambda _1\right) }\left( \omega
-1\right) -1.  \label{eq5-8}
\end{equation}
The fluctuations are relevant for 
\begin{equation}
\omega >-\frac{\ln C}{\ln \lambda _1},  \label{eq5-9}
\end{equation}
where the prefactor $C$ can be obtained from the linear analysis of the
unstable fixed point of the uniform ($J_A=J_B$) system. Although the
existence of some counterexample should not be ruled out, the validity of
this criterion has been confirmed by the application to a fair number of
cases (including the examples of Sections 3 and 4). In the limit of infinite
branching, the prefactor $C$ tends to unit, and the criterion of relevance
is reduced to the Pisot condition\cite{ref3}, $\omega >0$.

{\bf Acknowledgments}

We thank a suggestion of D. Mukamel and several discussions with R. F. S.
Andrade. STRP is on leave from Instituto de F\'{i}sica, Universidade Federal
da Bahia, and thanks a fellowship from the program CAPES/PICD. TASH and SRS
are partially supported by fellowships from CNPq.

\newpage\ 

\begin{center}
{\bf Figure Captions}
\end{center}

\bigskip\ 

{\bf Fig. 1-} Some stages of the construction of a diamond hierarchical
lattice with $q=2$ branches. Letters $A$ and $B$ indicate the ferromagnetic
exchange interactions ($J_A>0$ and $J_B>0$).

{\bf Fig. 2-} Basic graphs to obtain the recursion relations for an Ising
model on a diamond hierarchical lattice with $q=2$ and exchange interactions
given by the generalized Fibonacci rules, $A\rightarrow AB$, and $%
B\rightarrow AA$.

{\bf Fig. 3-} Second iterates and fixed points (black circles) of the
recursion relations in the parameter space, $x_A=\tanh K_A$ versus $%
x_B=\tanh K_B$, for an Ising model on a diamond hierarchical lattice with $%
q=2$ and exchange interactions given by the generalized Fibonacci rules, $%
A\rightarrow AB$, and $B\rightarrow AA$ (irrelevant fluctuations). We draw
the stable trivial fixed points (at zero and infinite temperatures) and the
physical saddle point. The dashed lines and the black arrows indicate the
flows of the second iterates of the map. The uniform model is recovered
along the diagonal, $x_A=x_B$. The light arrows indicate the stable
direction in the neighborhood of the physical fixed point.

{\bf Fig. 4-} Second iterates and fixed points for the Ising chain: (a)
exchange interactions according to the rules $A\rightarrow AB$, and $%
B\rightarrow AA$ (irrelevant fluctuations); (b) exchange interactions
according to the rules $A\rightarrow ABB$, and $B\rightarrow AAA$ (relevant
fluctuations). The diagonal solid line corresponds to the uniform model.

{\bf Fig. 5-} Second iterates and fixed points (black circles) of the
recursion relations in the parameter space, $x_A=\tanh K_A$ versus $%
x_B=\tanh K_B$, for an Ising model on a diamond hierarchical lattice with $%
q=2$ and exchange interactions given by the generalized Fibonacci rules, $%
A\rightarrow ABB$, and $B\rightarrow AAA$ (relevant fluctuations). We draw
the stable trivial fixed points (at zero and infinite temperatures) and the
unstable node. The uniform model is recovered along the diagonal, $x_A=x_B$.
The dashed lines are indications of the flow associated with the second
iterates of the map. The light arrows indicate the directions of the
eigenvectors in the neighborhood of the physical fixed point.

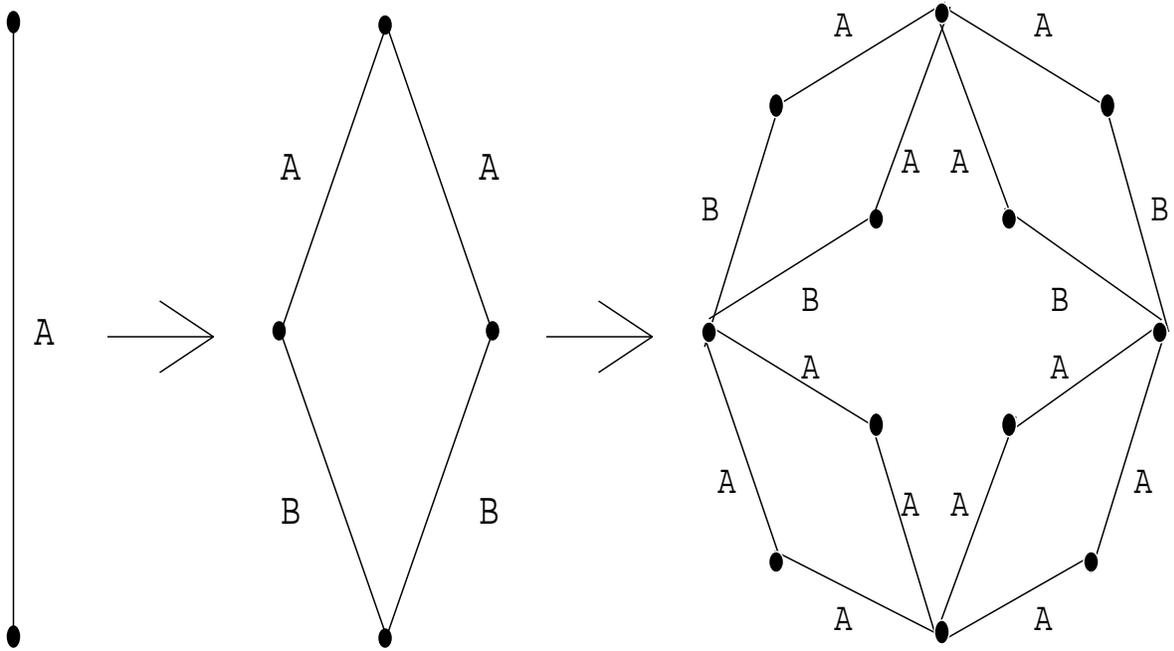
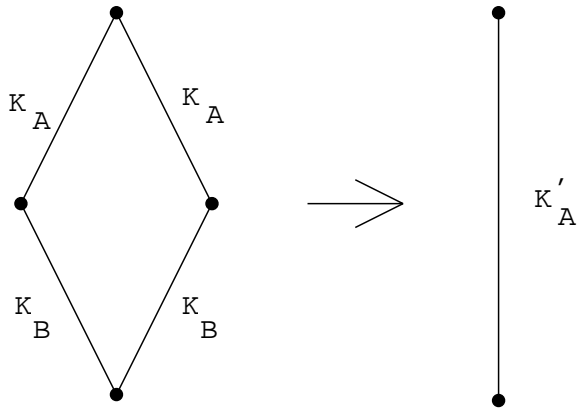
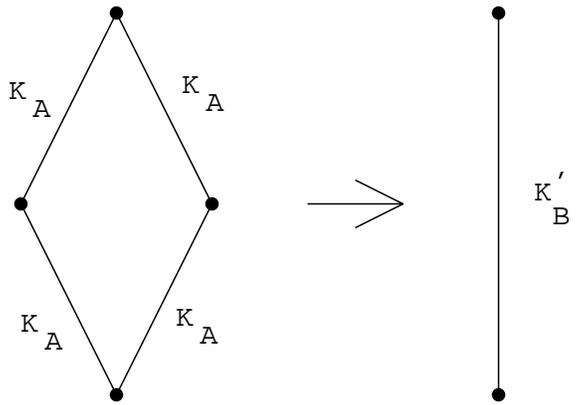
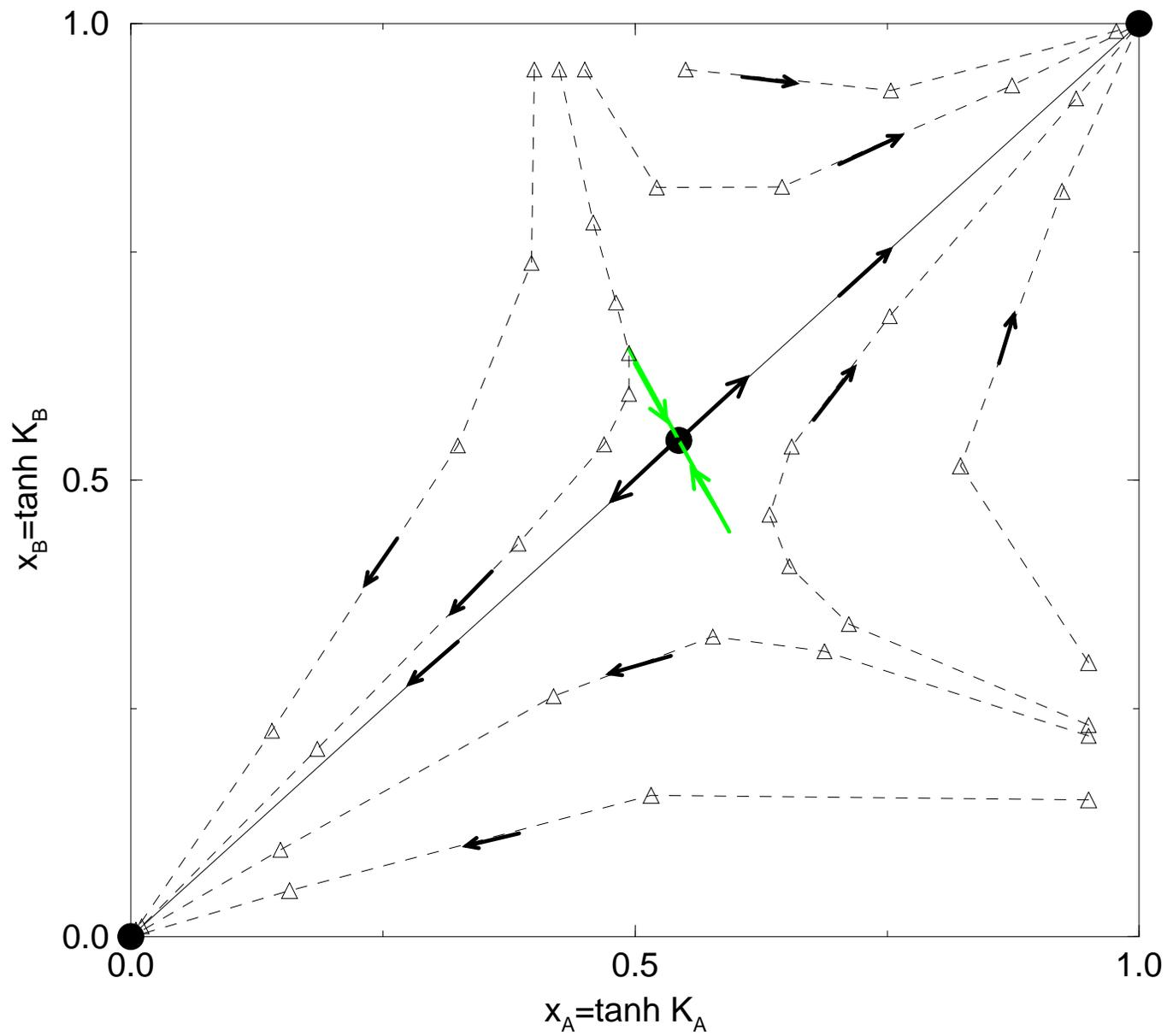
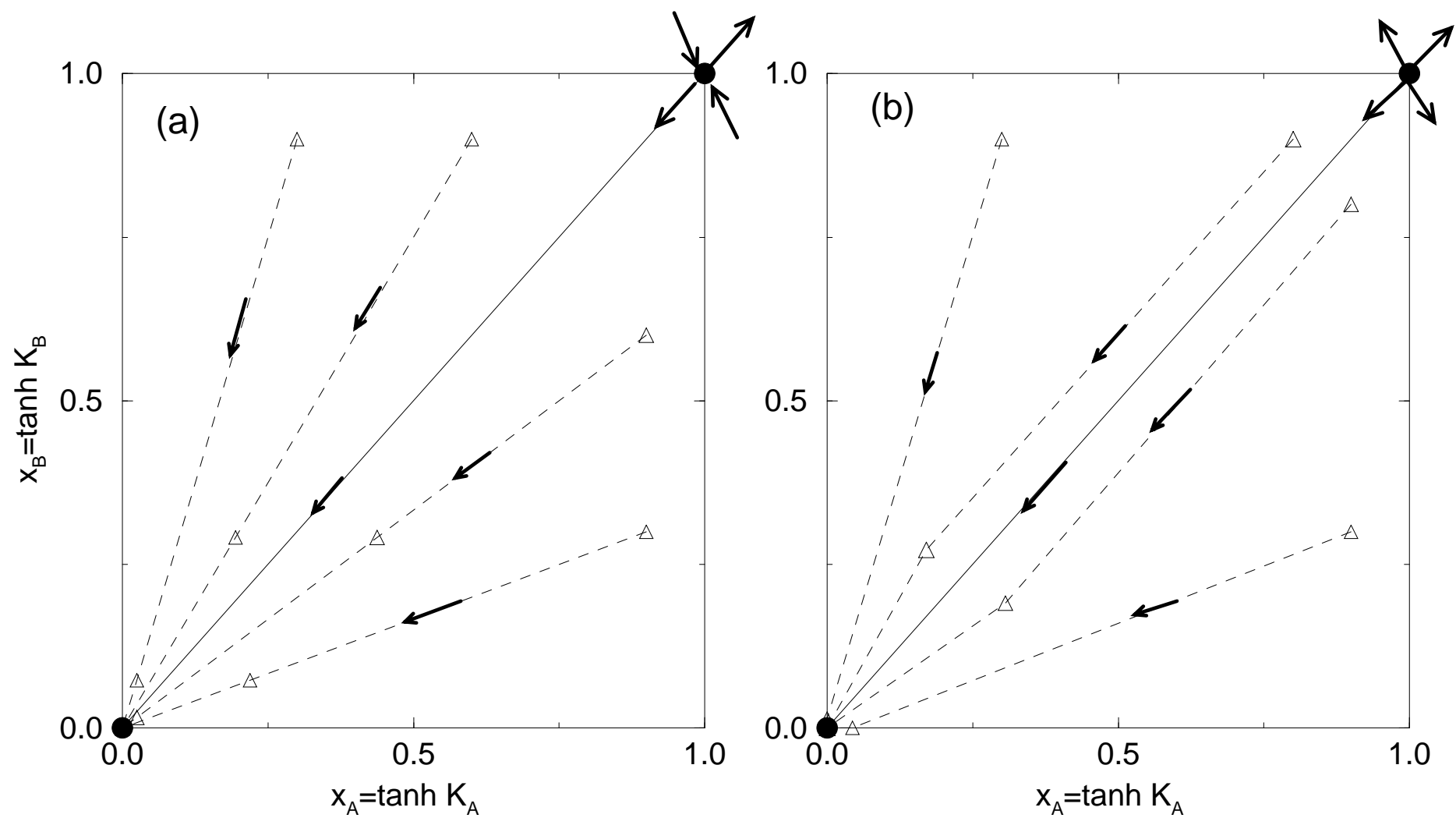
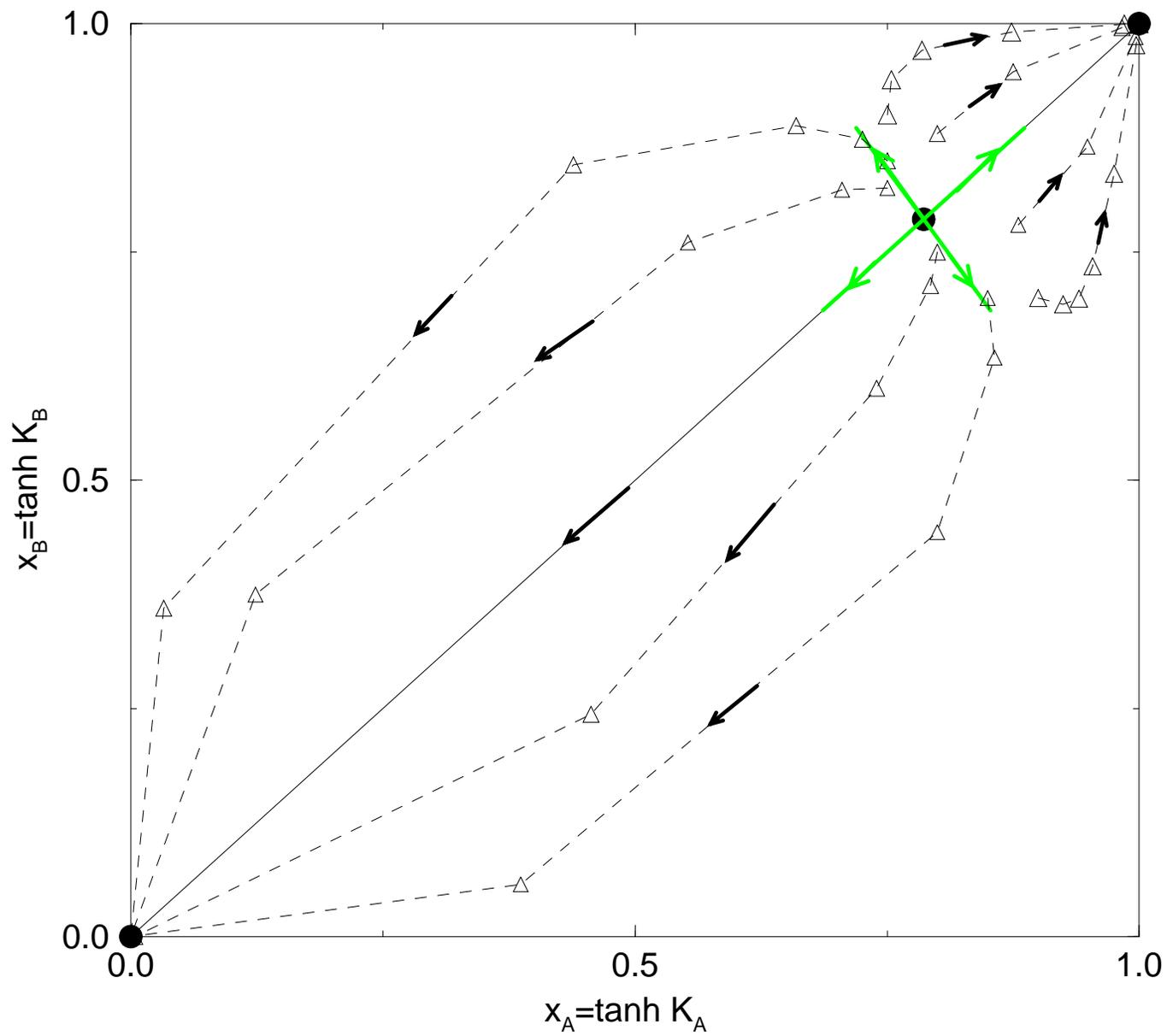
\end{document}